\documentclass[conference]{IEEEtran}
\usepackage[noadjust]{cite}
\usepackage[cmex10]{amsmath}
\usepackage{amsfonts}
\usepackage{amssymb}
\usepackage{algorithm}
\usepackage{algorithmic}
\usepackage[dvipsnames]{xcolor}
\usepackage[colorlinks, linkcolor=BrickRed, anchorcolor=blue, citecolor=blue, pdfstartview=FitH]{hyperref}
\usepackage{graphicx}
\usepackage{verbatim}
\usepackage{indentfirst}
\usepackage{booktabs}
\usepackage{epsfig}

\IEEEoverridecommandlockouts
\begin{document}

\title{Cognitive UAV Communication via Joint Trajectory and Power Control}
\IEEEspecialpapernotice{(Invited Paper)}
\author{Yuwei Huang$^1$, Jie Xu$^2$, Ling Qiu$^1$, and Rui Zhang$^3$\\
$^1$School of Information Science and Technology, University of Science and Technology of China\\
$^2$School of Information Engineering, Guangdong University of Technology\\
$^3$Department of Electrical and Computer Engineering, National University of Singapore\\
E-mail:~hyw1023@mail.ustc.edu.cn,~jiexu@gdut.edu.cn,~lqiu@ustc.edu.cn,~elezhang@nus.edu.sg}
\maketitle
\newcommand{\mv}[1]{\mbox{\boldmath{$ #1 $}}}
\newtheorem{lemma}{\underline{Lemma}}[section]
\newtheorem{remark}{\underline{Remark}}[section]
\setlength{\abovedisplayskip}{1pt}
\setlength{\belowdisplayskip}{1pt}
\begin{abstract}
This paper investigates a new spectrum sharing scenario between unmanned aerial vehicle (UAV) and terrestrial wireless communication systems. We consider that a cognitive/secondary UAV transmitter communicates with a ground secondary receiver (SR), in the presence of a number of primary terrestrial communication links that operate over the same frequency band. We exploit the UAV's controllable mobility via trajectory design, to improve the cognitive UAV communication performance while controlling the co-channel interference at each of the primary receivers (PRs). In particular, we maximize the average achievable rate from the UAV to the SR over a finite mission/communication period by jointly optimizing the UAV trajectory and transmit power allocation, subject to constraints on the UAV's maximum speed, initial/final locations, and average transmit power, as well as a set of interference temperature (IT) constraints imposed at each of the PRs for protecting their communications. However, the joint trajectory and power optimization problem is non-convex and thus difficult to be solved optimally. To tackle this problem, we propose an efficient algorithm that ensures to obtain a locally optimal solution by applying the techniques of alternating optimization and successive convex approximation (SCA). Numerical results show that our proposed joint UAV trajectory and power control scheme significantly enhances the achievable rate of the cognitive UAV communication system, as compared to benchmark schemes.
\end{abstract}

\begin{IEEEkeywords}
Unmanned aerial vehicle (UAV), UAV communication, cognitive radio, trajectory design, power control.
\end{IEEEkeywords}

\section{Introduction}
Unmanned aerial vehicles (UAVs) or drones are anticipated to have abundant civil applications in the future, for e.g. cargo delivery, agriculture inspection, surveillance, rescue and search, and communication relaying \cite{RZhangUAVchallenges}. As the number of UAVs increases explosively, it is crucial to provide them with seamless wireless data connections, in order to not only support secure, reliable, and low-latency remote command and control, but also enable high-capacity mission-related data transmission. There are generally two approaches to realize UAVs' communication with their ground users, namely the conventional direct UAV-to-ground communication and the newly proposed cellular-connected UAV communication \cite{RZhangUAVcommunication}.  In the former approach, UAVs are directly connected with ground control stations via point-to-point wireless communications; while in the latter case, UAVs are integrated into cellular networks as a new type of mobile users. As compared to the conventional direct UAV-to-ground communication, the cellular-connected UAV  can considerably improve the communication performance in terms of reliability, throughput, security, etc., and thus significantly increase the UAVs' operation range.

Due to the scarcity of wireless spectrum, for both approaches above, UAVs may need to share the spectrum with existing wireless devices (e.g., cellular mobiles on the ground) for communications \cite{WZspectrumsharing}. This resembles spectrum sharing in cognitive radio (CR) networks, in which secondary users share the same frequency bands with existing primary users \cite{GoldSmith}. In this case, the UAV-to-ground communication may cause severe interference to the existing terrestrial users, as UAVs usually have strong line-of-sight (LoS) links with ground nodes such as cellular base stations (BSs), due to their high altitude over the air. As a result, how to optimize the UAV communication performance while effectively controlling the air-to-ground co-channel interference is a new and challenging problem to be solved. By leveraging the UAV's controllable mobility, in this paper, we propose a new approach to solve this problem, which jointly optimizes the UAV trajectory and transmit power allocation to achieve the maximum throughput of the UAV-to-ground communication and yet control the interference to existing ground receivers below a tolerable level.

Specifically, this paper considers a cognitive UAV communication system, where a cognitive/secondary UAV transmitter communicates with a ground secondary receiver (SR), in the presence of a number of primary terrestrial communication links that operate over the same frequency band. We adopt the {\it interference temperature} (IT) method in CR networks \cite{RZhangIT}, \cite{RZhangIT2} to protect the primary communication links, based on which the received interference power at each primary receiver (PR) cannot exceed a prescribed IT threshold. Under this setup, we maximize the average achievable rate of the cognitive UAV communication  over a finite UAV mission/communication period, by jointly optimizing the UAV trajectory and transmit power allocation, subject to the maximum speed, initial/final locations and average transmit power constraints of the UAV, as well as the average IT constraints at the PRs.

However, the joint trajectory and power optimization problem is non-convex and thus difficult to be solved optimally. To tackle this problem, we propose an efficient algorithm that ensures to obtain a locally optimal solution by applying the techniques of alternating optimization and successive convex approximation (SCA). Numerical results show that our proposed joint UAV trajectory and power control scheme significantly improves the achievable rate of the cognitive UAV communication system, as compared to benchmark schemes with trajectory optimization or power control only.

Note that in the literature, there have been a handful of works that studied the UAV's trajectory design for improving the UAV communication performance under different setups \cite{RZhangUAVcommunication}, \cite{RZhangrelay}--\cite{RZhangUAVcommunication2}. For example,  \cite{RZhangrelay}, \cite{DGrelay} employed the UAV as a mobile relay to help enhance the communication throughput between two ground users. \cite{RZhangbs}--\cite{RZhangUAVcommunication2} employed UAVs as aerial BSs to broadcast individual information or multicast common information to  a set of ground users. A cellular-connected UAV application was considered in \cite{RZhangUAVcommunication}, which optimized the UAV trajectory to minimize the mission completion time, subject to the communication connectivity constraints with ground BSs.  Furthermore, in another line of work, UAVs were considered as mobile access points (APs) for charging ground Internet-of-things (IoT) devices \cite{JXuWPT} and simultaneously collecting information from them \cite{JXu2018}. Different from these prior works, this paper aims to investigate the new spectrum sharing scenario between UAV and terrestrial wireless communication systems, while we exploit the joint UAV trajectory design and transmit power control for both enhancing the UAV communication throughput as well as effectively controlling the air-to-ground interference to terrestrial users.

\section{System Model}
In this paper, we consider the scenario where a cognitive/secondary UAV  transmitter sends information to a ground SR, in the presence of a set of $K\ge 1$ primary users that operate over the same frequency band.  Let $\mathcal{K}\triangleq \{1,\ldots,K\}$ denote the set of ground PRs. This may correspond to the uplink transmission from the UAV to its associated ground BS (SR) in a cellular network, while there are $K$ ground users in the neighborhood simultaneously transmitting to their respective ground BSs (PRs) at the same frequency band. We focus on the cognitive UAV communication over a particular mission period, denoted by $\mathcal{T}=[0,T]$, with duration $T>0$ in second~(s).

We consider a three-dimensional (3D) Cartesian coordinate system, where the SR and each PR $k\in \mathcal K$ have fixed locations of $\mv{w}=(x,y)$ and $\mv{w}_{k}=(x_{k},y_{k})$, respectively. It is assumed that the UAV perfectly knows the locations of the ground SR and PRs {\it a-priori} to facilitate the joint trajectory and power control design. We assume that the UAV flies at a constant altitude $H>0$ in meter (m) with the time-varying horizontal location $\hat{\mv{{q}}}(t)=(\hat{x}(t),\hat{y}(t))$, $t\in\mathcal T$. Specifically, the UAV's initial and final (horizontal) locations are pre-determined as $\hat{\mv q}_{I}=(x_{I},y_{I})$ and $\hat{\mv q}_{F}=(x_{F},y_{F})$, respectively. Let $\hat{V}$ denote the maximum UAV speed in m/s. Then we have $\sqrt{\dot{\hat{x}}^{2}(t)+\dot{\hat{y}}^{2}(t)}\le \hat{V}$, $\forall t\in\mathcal T$, where $\dot{\hat{x}}(t)$ and $\dot{\hat{y}}(t)$ denote the first derivatives of $\hat{x}(t)$ and $\hat{y}(t)$, respectively. For ease of exposition, we discretize the mission/communication period $\mathcal T$ into $N$ time slots each with equal duration $\delta_{t}=T/N$, where $N$ is chosen to be sufficiently large such that the UAV location can be assumed to be approximately constant within each time slot. Accordingly, let $\mv{q}[n]=(x[n],y[n])$ denote the horizontal UAV location at time slot $n\in\mathcal N \triangleq \{1,\ldots,N\}$. As a result, we have the following constraints on the UAV trajectory.
\begin{align}
\Vert\mv{q}[n]-\mv{q}[n-1]\Vert^{2}&\leq V^{2},\label{UAV trajectory1}\\
\mv{q}[0]&=\hat{\mv{q}}_{I},\label{UAV trajectory2}\\
\mv{q}[N]&=\hat{\mv{q}}_{F},\label{UAV trajectory3}
\end{align}
where $V\triangleq \hat{V}\delta_{t}$ denotes the maximum UAV displacement during each time slot, and $\Vert\cdot\Vert$ denotes the Euclidean norm. Furthermore, at time slot $n\in\mathcal N$, the distance between the UAV and the SR and that between the UAV and each PR $k\in\mathcal K$ are respectively given by
\begin{align}
d(\mv q[n])&=\sqrt{H^{2}+\Vert \mv{q}[n]-\mv{w}\Vert^{2}},\\
d_{k}(\mv q[n])&=\sqrt{H^{2}+\Vert \mv{q}[n]-\mv{w}_{k}\Vert^{2}}.
\end{align}
~~In practice, the air-to-ground wireless channels are normally dominated by the LoS link \cite{RZhangrelay}. Therefore, similarly as in \cite{RZhangrelay}, we consider the free-space path-loss model for the wireless channels from the UAV to the SR and PRs. As a result, at time slot $n\in\mathcal N$, the channel power gain from the UAV to the SR and that to each PR $k\in\mathcal K$ are respectively expressed as
\begin{align}
h(\mv q[n])&=\beta_{0}d^{-2}(\mv q[n])=\frac{\beta_{0}}{H^{2}+\Vert \mv{q}[n]-\mv{w}\Vert^{2}} ,\\
g_{k}(\mv q[n])&=\beta_{0}d_{k}^{-2}(\mv q[n])=\frac{\beta_{0}}{H^{2}+\Vert \mv{q}[n]-\mv{w}_{k}\Vert^{2}} ,
\end{align}
where $\beta_{0}$ denotes the channel power gain at the reference distance of $d_{0}=1\ \text{m}$.
Accordingly, by letting $p[n]\geq 0$ denote the transmit power of the UAV at time slot $n\in\mathcal N$, the achievable rate from the UAV to the SR in bits/second/Hertz (bps/Hz) at time slot $n$ is
\begin{align}
R\left(p[n],\mv q[n]\right)&=\log_{2}\left(1+\frac{h(\mv q[n])p[n]}{\sigma^{2}}\right),\nonumber\\
&=\log_{2}\left(1+\frac{\eta_{0}p[n]}{H^{2}+\Vert \mv{q}[n]-\mv{w}\Vert^{2}}\right),
\end{align}
where $\sigma^{2}$ denotes the noise power at the SR receiver, and $\eta_{0}=\beta_{0}/\sigma^{2}$ denotes the reference signal-to-noise ratio (SNR). Note that $\sigma^{2}$ also takes into account the interference from the primary transmitters (PTs). Let $P$ denote the maximum average transmit power at the UAV. We thus have
\begin{align}\label{pave}
\frac{1}{N}\sum\limits_{n=1}^{N}p[n]\leq P.
\end{align}
~~Under spectrum sharing, the secondary UAV communication system introduces air-to-ground co-channel interference to the ground PRs. At time slot $n\in\mathcal N$, the interference power from the UAV to each PR $k\in\mathcal K$ is
\begin{align}
Q_{k}\left(p[n],\mv q[n]\right)&=g_{k}(\mv q[n])p[n]=\frac{\beta_{0}p[n]}{H^{2}+\Vert \mv{q}[n]-\mv{w}_{k}\Vert^{2}}.
\end{align}In order to protect the primary communications, we apply the IT constraint \cite{RZhangIT}, \cite{RZhangIT2} at each PR $k$, such that the received average interference power does not exceed the IT threshold, denoted by $\Gamma_{k}\geq 0,\ k\in\mathcal K$.\footnote{In this work, we consider the average IT constraint instead of the peak IT constraint, as it has been shown in \cite{RZhangIT2} that the former leads to better achievable rates than the latter for both the primary and secondary links, under the same total resulted interference power over time.} We thus have
\begin{align}\label{ITave}
\frac{1}{N}\sum\limits_{n=1}^{N}\frac{\beta_{0}p[n]}{H^{2}+\Vert \mv{q}[n]-\mv{w}_{k}\Vert^{2}}\leq \Gamma_{k},\ \forall k\in \mathcal K.
\end{align}

Our objective is to maximize the average achievable rate of the secondary UAV communication system (i.e., $\small\frac{1}{N}\sum\limits_{n=1}^{N}R\left(p[n],\mv q[n]\right)$), by jointly optimizing the UAV trajectory $\{\mv q[n]\}$ and the transmit power allocation $\{p[n]\}$, subject to the UAV maximum speed constraint in (\ref{UAV trajectory1}), the initial/final location constraints in (\ref{UAV trajectory2}) and (\ref{UAV trajectory3}), the average transmit power constraint in (\ref{pave}), and the IT constraints in (\ref{ITave}). Therefore, the problem of our interest is formulated as
\begin{align}
(\text{P1}):\ \max\limits_{\left\{p[n],\mv{q}[n]\right\}}&\frac{1}{N}\sum\limits_{n=1}^{N}\log_{2}\left(1+\frac{\eta_{0}p[n]}{H^{2}+\Vert\mv{q}[n]-\mv{w}\Vert^{2}}\right)\nonumber \\
\text{s.t.}\ &p[n]\geq 0,\ \forall n\in \mathcal N,\label{p0}\\
&\text{(\ref{UAV trajectory1}),~(\ref{UAV trajectory2}),~(\ref{UAV trajectory3}),~(\ref{pave})},\ \text{and}\ \text{(\ref{ITave})}.\nonumber
\end{align}
Note that problem (P1) is a non-convex optimization problem, as the objective function is non-concave and the constraints in (\ref{ITave}) are non-convex. Therefore, this problem is generally difficult to be solved optimally.
\begin{remark}
It is worth nothing that under given UAV trajectory $\{\mv q[n]\}$, the transmit power allocation in (P1)  is reminiscent of that for throughput maximization in fading CR channels (see, e.g., \cite{XKangpowerfadingchannel}). However, different from conventional fading CR channels with random wireless channel fluctuations, the cognitive UAV communication system can properly design the UAV trajectory for controlling the wireless channel power gains over time (see (P1)). This thus provides a new and unique design degree of freedom for communication performance optimization.
\end{remark}

\vspace{-1em}
\section{Proposed Solution to Problem (P1)}
\vspace{-0.5em}
In this section, we present an efficient algorithm based on alternating optimization, to obtain a locally optimal solution to (P1), by optimizing one of the transmit power $\{p[n]\}$ and the UAV trajectory $\{\mv q[n]\}$ with the other fixed in an alternating manner.
\vspace{-0.5em}
\subsection{Transmit Power Optimization Under Given Trajectory}
First, we optimize the transmit power allocation $\{p[n]\}$ under any given UAV trajectory $\{\mv{q}[n]\}$, for which the problem is expressed as
\begin{align}
(\text{P2}):\ \max\limits_{\{p[n]\}}&\frac{1}{N}\sum\limits_{n=1}^{N}\log_{2}\left(1+\frac{\eta_{0}p[n]}{H^{2}+\Vert\mv{q}[n]-\mv{w}\Vert^{2}}\right)\nonumber\\
\text{s.t.}\ &\text{(\ref{pave}),\ (\ref{ITave})},\ \text{and}\ \text{(\ref{p0})}.\nonumber
\end{align}
Notice that problem (P2) is a convex optimization problem, as the objective function of (P2) is concave with respect to $\{p[n]\}$, and all the constraints are convex. Therefore, problem (P2) can be solved optimally by standard convex optimization techniques, such as the interior point method \cite{Convex}.
\vspace{-0.5em}
\subsection{Trajectory Optimization Under Given Transmit Power}
Next, we optimize the UAV trajectory $\{\mv{q}[n]\}$ under any given transmit power $\{p[n]\}$, for which the problem is formulated as
\begin{align}
(\text{P3}):\ \max\limits_{\left\{\mv{q}[n]\right\}}&\frac{1}{N}\sum\limits_{n=1}^{N}\log_{2}\left(1+\frac{p[n]\eta_{0}}{\Vert\mv{q}[n]-\mv{w}\Vert^{2}}\right)\nonumber\\
\text{s.t.}\ &\text{(\ref{UAV trajectory1}),~(\ref{UAV trajectory2}),~(\ref{UAV trajectory3})},\ \text{and}\ \text{(\ref{ITave})}.\nonumber
\end{align}
Notice that problem (P3) is non-convex, as the objective function is non-concave with respect to $\mv q[n]$ and the constraints in (\ref{ITave}) are non-convex. To tackle this problem, we adopt the SCA technique to obtain a locally optimal solution to (P3) in an iterative manner. The key idea of the SCA is that given a local point at each iteration, we approximate the non-concave objective function (or the non-convex constraints) into a concave objective function (convex constraints), in order to obtain an approximated convex optimization problem. By iteratively solving a sequence of approximated convex problems, we can obtain an efficient solution to the original non-convex optimization problem (P3).

Specifically, suppose that $\{\mv q^{(j)}[n]\}$ corresponds to the obtained UAV trajectory at the $(j-1)$-th iteration with $j\geq 1$, where $\{\mv q^{(0)}[n]\}$ corresponds to the initial UAV trajectory. In the following, we explain how to approximate the objective function of (P3) and the constraints in (\ref{ITave}), respectively. First, as for the non-concave objective function of (P3), we have the following lemma.
\begin{lemma}\label{SCAl}
For any given $\{\mv q^{(j)}[n]\}$, $j\ge 0$, it follows that
\begin{align}\label{SCAlb}
R\left(p[n],\mv q[n]\right)\geq R^{\text{lb}}\left(p[n],\mv q[n]\right),
\end{align}
where

\vspace{-1em}
\begin{small}
\begin{align}
&R^{\text{lb}}\left(p[n],\mv q[n]\right)\triangleq \log_{2}\left(1+\frac{\eta_{0}p[n]}{H^{2}+\Vert \mv{q}^{(j)}[n]-\mv{w}\Vert^{2}}\right)\nonumber\\
&-\frac{\eta_{0}p[n]\log_{2}e\left(\Vert\mv{q}[n]-\mv{w}\Vert^{2}-\Vert\mv{q}^{(j)}[n]-\mv{w}\Vert^{2}\right)}{\left(H^{2}+\Vert\mv q[n]-\mv w\Vert^{2}\right)\left(\left(H^{2}+\Vert\mv q[n]-\mv w\Vert^{2}\right)+\eta_{0}p[n]\right)},
\end{align}
\end{small}and the inequality in (\ref{SCAlb}) is tight for $\mv{q}[n]=\mv{q}^{(j)}[n]$.
\end{lemma}
\begin{IEEEproof}
By introducing an auxiliary variable $\alpha=\Vert\mv q[n]-\mv w\Vert^{2}\geq 0$, we have $R(p[n],\mv q[n])=\tilde R(p[n],\alpha)\triangleq\log_{2}\left(1+\frac{\eta_{0}p[n]}{H^{2}+\alpha}\right)$. It is evident that $\tilde R(p[n],\alpha)$ is a convex function with respect to $\alpha\geq 0$. Therefore, $\tilde R(p[n],\alpha)$ can be globally lower-bounded by its first-order Taylor expansion with respect to $\alpha$ at any point. By doing so and substituting $\alpha=\Vert \mv q[n]-\mv w\Vert^{2}$, this lemma is proved.
\end{IEEEproof}

Next, consider the non-convex constraints in (\ref{ITave}), which can be equivalently expressed as the following constraints by introducing auxiliary variables $\{t_{k}[n]\}$.
\begin{align}
t_{k}[n]&\leq\Vert\mv{q}[n]-\mv{w}_{k}\Vert^{2},\ \forall n\in\mathcal N,~k\in\mathcal K,\label{tk}\\
t_{k}[n]&\geq 0,\ \forall n\in\mathcal N,~k\in\mathcal K,\label{t0}\\
\frac{1}{N}\sum\limits_{n=1}^{N}\frac{\beta_{0}p[n]}{t_{k}[n]}&\leq \Gamma_{k},\ \forall k\in\mathcal K.\label{SCAIT}
\end{align}
Notice that the constraints in (\ref{t0}) and (\ref{SCAIT}) are both convex, while only those in (\ref{tk}) are still non-convex. Since $\Vert\mv{q}[n]-\mv{w}_{k}\Vert^{2}$ is a convex function with respect to $\mv{q}[n]$, we have the following inequalities by applying the first-order Taylor expansion at any given point $\{\mv{q}^{(j)}[n]\}$:

\vspace{-1em}
\begin{small}
\begin{align}
&\Vert\mv{q}[n]-\mv{w}_{k}\Vert^{2}\geq\Vert\mv{q}^{(j)}[n]-\mv{w}_{k}\Vert^{2}\nonumber\\
&+2\left(\mv{q}^{(j)}[n]-\mv{w}_{k}\right)^{T}\left(\mv{q}[n]-\mv{q}^{(j)}[n]\right),\ \forall n\in\mathcal N,~k\in\mathcal K.
\label{SCA2}\end{align}
\end{small}By replacing $\Vert \mv q[n]-\mv w_{k}\Vert^{2}$ in (\ref{tk}) as the right-hand-side (RHS) of (\ref{SCA2}), we approximate (\ref{tk}) as the following convex constraints:

\vspace{-1em}
\begin{small}
\begin{align}
&t_{k}[n]\leq\Vert\mv{q}^{(j)}[n]-\mv{w}_{k}\Vert^{2}\nonumber\\
&+2\left(\mv{q}^{(j)}[n]-\mv{w}_{k}\right)^{T}\left(\mv{q}[n]-\mv{q}^{(j)}[n]\right),\forall n\in\mathcal N,~k\in \mathcal K.\label{tkq}
\end{align}
\end{small}To summarize, by replacing $R\left(p[n],\mv q[n]\right)$ in the objective function as $R^{\text{lb}}\left(p[n],\mv q[n]\right)$, and replacing the constraints in (\ref{ITave}) as those in (\ref{t0}), (\ref{SCAIT}), and (\ref{tkq}), problem (P3) is approximated as the following convex optimization problem (P3.1) at any local point $\{\mv q^{(j)}[n]\}$, which can be solved via standard convex optimization techniques such as the interior point method \cite{Convex}, with the optimal solution denoted as $\{\mv q^{(j)*}[n]\}$ and $\{t_{k}^{(j)*}[n]\}$.
\begin{align}
(\text{P3.1}):\ \max\limits_{\{\mv{q}[n],t_{k}[n]\}}&\frac{1}{N}\sum\limits_{n=1}^{N}R^{\text{lb}}\left(p[n],\mv q[n]\right)\nonumber\\
\text{s.t.}\ &(\ref{UAV trajectory1}),~(\ref{UAV trajectory2}),~(\ref{UAV trajectory3}),~(\ref{t0}),~(\ref{SCAIT}),~\text{and}~(\ref{tkq}).\nonumber
\end{align}
With the convex optimization problem (P3.1) at hand, we can obtain an efficient algorithm to solve (P3) in an iterative manner. In the $j$-th iteration, this algorithm solves the convex optimization problem (P3.1) at the local point $\{\mv q^{(j)}[n]\}$, where $\{\mv q^{(j)}[n]\}$ corresponds to the optimal trajectory solution to (P3.1) obtained in the previous iteration $(j-1)$, i.e., $\mv q^{(j)}[n]=\mv q^{(j-1)*}[n]$. We summarize this algorithm in Table \ref{SCA} as Algorithm 1.
\vspace{-2em}
\begin{table}[!htb]\scriptsize
\caption{Algorithm 1 for Solving Problem (P3)} \centering
\vspace{-1.2em}
\begin{tabular}{|p{8.5cm}|}
\hline\vspace{-1em}
\begin{itemize}
\item[a)]{\bf Initialization:} Set the initial UAV trajectory as $\{\mv q^{(0)}[n]\}_{n=1}^{N}$, and $j=0$.
\item[b)] {\bf Repeat:}
\begin{itemize}
\item[1)] Solve problem (P3.1) to obtain the optimal solution as $\{\mv q^{(j)*}[n]\}_{n=1}^{N}$ and $\{t_{k}^{(j)*}[n]\}_{n=1}^{N}$.
\item[2)] Update the trajectory as $\mv q^{(j+1)}[n]=\mv q^{(j)*}[n]$, $\forall n\in \mathcal N$.
\item[3)] Update $j=j+1$.
\end{itemize}
\item[c)] {\bf Until} the objective value of (P3) converges within a given accuracy or a maximum number of iterations is reached.
\end{itemize}\vspace{-1em}\\ \hline
\end{tabular}\label{SCA}\vspace{-1em}
\end{table}

\vspace{-0.4em}
It is easy to show that in Algorithm 1, after each iteration $j$, the objective function of (P3) achieved by $\{\mv q^{(j)}[n]\}$ is monotonically non-decreasing \cite{JXuWPT}. As the optimal value of problem (P3) is upper-bounded, it is evident that Algorithm 1 can converge to a locally optimal solution to problem (P3).
\vspace{-0.5em}
\subsection{Alternating Optimization}
\vspace{-0.4em}
Now, we are ready to present a complete algorithm to solve (P1) via alternating optimization. This algorithm optimizes the transmit power $\{p[n]\}$ by solving (P2) under given UAV trajectory $\{\mv q[n]\}$, as well as the trajectory $\{\mv q[n]\}$ with given transmit power $\{p[n]\}$ by solving (P3) via Algorithm 1, in an alternating manner. Notice that at each iteration, the algorithm ensures that the objective value of (P1) is monotonically non-decreasing. As the optimal value of (P1) is upper-bounded, the alternating optimization algorithm is ensured to coverage to a locally optimal solution to (P1).
\section{Numerical Results}
In this section, we present numerical results to validate the performance of our proposed design with joint UAV trajectory and power optimization. We set the maximum UAV speed as $\hat{V}=50~\text{m/s}$, the noise power at the SR as $\sigma^{2}=-50~\text{dBm}$, the channel power gain at the reference distance of 1 m as $\beta_{0}=-30~\text{dB}$, the average transmit power as $P=30~\text{dBm}$, and the UAV's fixed flight altitude as $H=100~\text{m}$. Furthermore, we consider that the SR has the horizontal location $(0~\text{m},~0~\text{m})$, and there are two PRs with horizontal locations $(-500~\text{m},~500~\text{m})$ and $(500~\text{m},\ -500~\text{m})$, respectively. The UAV's initial and final horizontal locations are set as $(-1000~\text{m},~1000~\text{m})$ and $(1000~\text{m},~-1000~\text{m})$, respectively, and the IT constraints are identical for different PRs, i.e., $\Gamma_{k}=\Gamma,~\forall k\in\mathcal K$. In addition, for Algorithm 1, we choose the initial UAV trajectory following a straight line, in which the UAV flies directly from the initial location to the final location with a constant speed $\tilde{V}=\Vert\hat{\mv{q}}_{F}-\hat{\mv{q}}_{I}\Vert/T$, which is less than the maximum speed $\hat{V}$ assumed.

\vspace{-1.3em}
\begin{figure}[!htb]
\centering
\epsfxsize=1\linewidth
    \includegraphics[width=5cm]{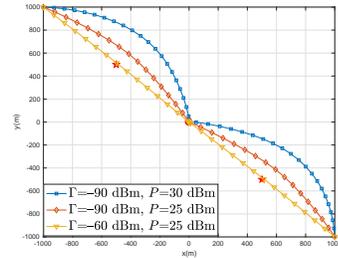}
    \vspace{-1.5em}
\caption{UAV trajectories projected onto the ground (horizontal) plane by the proposed design with joint UAV trajectory and power optimization. The red stars represent the locations of the two ground PRs, respectively, and the red circle denotes the location of the ground SR.}
\label{tra}
\end{figure}
\vspace{-1.2em}
Fig.~\ref{tra} shows the obtained UAV trajectories by the proposed design with joint UAV trajectory and power optimization, under different values of average transmit power constraint $P$ and average IT constraint $\Gamma$, where the communication/mission duration is set as $T=200~\text{s}$. Note that the trajectories shown are projected onto the ground (horizontal) plane, and the points on each trajectory are sampled every 1 s. It is observed that when $\Gamma=-60~\text{dBm}$ and $P=25~\text{dBm}$, the UAV trajectory follows a straight line from the initial to the final location; when $\Gamma$ decreases (i.e., $\Gamma=-90~\text{dBm}$, and $P=25~\text{dBm}$), the UAV trajectory deviates from the straight line to move away from the PRs for minimizing the air-to-ground interference to them; when $P$ further increases (i.e., $\Gamma=-90~\text{dBm}$, and $P=30~\text{dBm}$), the UAV moves further away from the PRs.  It is also observed that for all the three trajectories, the sampled points become closer when the UAV moves near the SR, while they become further apart when the UAV is near each of the PRs. This indicates that the UAV flies above the SR with low or even zero speed for taking advantage of the best communication channel for transmission, but moves away from the PRs with high or even maximum speed for co-channel interference power minimization. Such UAV trajectories are intuitive, which show the benefit of mobility control in balancing the tradeoff between communication throughput maximization and co-channel interference minimization.

\begin{figure}[!t]
\setlength{\belowcaptionskip}{-3em} 
\centering
\epsfxsize=1\linewidth
    \includegraphics[width=5cm]{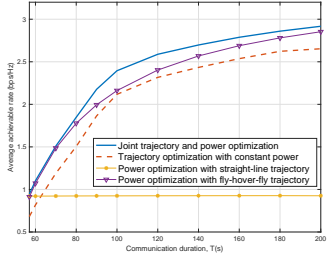}
    \vspace{-1.5em}
\caption{The average achievable rate of the cognitive UAV communication versus the communication duration $T$.}
\label{rave}
\end{figure}
Fig. \ref{rave} shows the average achievable rate of the cognitive UAV communication system versus the communication duration $T$, where we set $\Gamma=-60\ \text{dBm}$ and $P=30\ \text{dBm}$. For performance comparison, we also consider the following three benchmark schemes:
\begin{itemize}
\item[$\bullet$] {\bf Trajectory optimization with constant power}: The UAV optimizes its trajectory $\{\mv q[n]\}$ via Algorithm 1, where the transmit power is fixed as $p[n]=p,~\forall n\in\mathcal N$. Here, $p \ge 0$ is chosen as the maximum value such that the transmit power constraint $p \le P$ and the IT constraints at PRs are both satisfied. Under our setup in the simulation, we set $p = P$. 
\item[$\bullet$] {\bf Power optimization with straight-line trajectory}: The UAV sets its trajectory following a straight line from the initial to the final location with a constant speed. Under this trajectory, the UAV optimizes its power allocation by solving problem (P2).
\item[$\bullet$] {\bf Power optimization with fly-hover-fly trajectory}: The UAV first flies directly from the initial location to the location above SR at the maximum speed, then hovers above the SR for a certain (maximum) amount of time, and finally flies directly to the final location at the maximum speed. Under this trajectory, the UAV optimizes its power allocation by solving problem (P2).
\end{itemize}In Fig. \ref{rave}, it is observed that as the communication duration $T$ increases, the average achievable rate by the power optimization with straight-line trajectory remains unchanged, while those by the other three schemes increase. This is due to the fact that under the straight-line trajectory with constant UAV speed, the UAV has the same channel gain distribution with the SR (or each of the PRs), which is regardless of $T$. By contrast, for the other cases with adaptive trajectory design with $T$, the UAV in general stays longer near the SR when $T$ increases, thus leading to a better channel condition on average and thus a higher average achievable rate. When $T$ is small (e.g., $T\leq 60~\text{s}$), it is observed that the three schemes with power optimization outperform the trajectory optimization with constant power. This is because when $T$ is small, the gain of trajectory design cannot be fully exploited, and thus power optimization plays a more important role. By contrast, when $T$ becomes large (e.g., $T\geq 70~\text{s}$), the schemes with trajectory optimization are observed to outperform the power optimization with straight-line trajectory. This shows that trajectory optimization becomes more significant in this regime. Over all regimes, the proposed joint trajectory and power control design is observed to outperform the three benchmark schemes. This validates the practical throughput gain of such a joint optimization approach.

\section{Conclusion}
This paper studied a new spectrum sharing scenario, where a cognitive/secondary UAV communication system coexists with primary terrestrial wireless communication links. We optimized the UAV's trajectory, jointly with its transmit power allocation, to maximize the average achievable rate of the cognitive UAV communication system over a finite mission/communication period, subject to a set of IT constraints for protecting the PRs. To tackle this non-convex optimization problem, we proposed an efficient algorithm to obtain a locally optimal solution via alternating optimization and SCA. Numerical results validated the superior performance of our proposed design against other benchmark schemes.

\end{document}